\documentclass[preprint]{aastex62}

\shorttitle{X-rays and cosmic rays in Arches cloud}

\newcommand{\nustar}{{\it NuSTAR}}

\begin{document}

\title{Time-Variability of equivalent width of 6.4 keV line from the Arches Complex: reflected X-rays or charged particles?}

\author[0000-0003-0716-5951]{D. O. Chernyshov}\altaffiliation{chernyshov@dgap.mipt.ru}
\affiliation{I.~E.~Tamm Theoretical Physics Division of P.~N.~Lebedev Institute of Physics, 119991 Moscow, Russia}
\author{C. M. Ko}\altaffiliation{cmko@astro.ncu.edu.tw}
\affiliation{Institute of Astronomy, Department of Physics and Center for Complex Systems,
National Central University, Jhongli District, Taoyuan City, Taiwan 320, R.O.C.}
\author{R. A. Krivonos}
\affiliation{Space Research Institute of the Russian Academy of Sciences, Profsoyuznaya Str. 84/32, 117997 Moscow, Russia}
\author{V. A. Dogiel}
\affiliation{I.~E.~Tamm Theoretical Physics Division of P.~N.~Lebedev Institute of Physics, 119991 Moscow, Russia}
\author{K. S. Cheng}
\affiliation{Department of Physics, University of Hong Kong, Pokfulam Road, Hong Kong, China}

%\correspondingauthor{C.-M. Ko}
%\email{cmko@gm.astro.ncu.edu.tw}

\begin{abstract}
Molecular gas in Arches cloud located near the Arches cluster is one of the emitters of K-$\alpha$ line of neutral iron
and X-ray continuum in the Galactic center (GC). Similarly to the cloud Sgr B2, another well-known emitter of the iron line in the GC,
the Arches cloud demonstrates  temporal decline of the X-ray emission. The most natural origin of this emission is irradiation	
of primary photons of an X-ray flare from a distant source, most likely Sgr A$^*$. However, recent observations of the Arches cloud 		
discovered variations of equivalent width of the 6.4 keV iron line, which indicated that the X-ray emission from the cloud is 		
a combination of two components with different origin and different equivalent width, one of which is time-variable, 		
while the other is stationary during the period of observations. We considered two different scenarios: a)  		
this emission is formed by reflection from two  clouds, which are at some distance from each other, 		
when they are irradiated by two different flares; and b) the other scenario assumes a combination of X-ray fluxes produced
in the same cloud by reflection of primary photons and by subrelativistic cosmic rays.
We present restrictions for both model and conditions at which these scenarios can be realized.
Although none of the models can be completely ruled out, we find that the X-ray reflection model requires less
assumption and therefore is the most viable.
\end{abstract}


\section{Introduction}

X-ray emission from molecular clouds was detected by the GRANAT team in 1993 \citep[see][]{suny}.
They assumed that this emission was the Compton echo from molecular clouds which reflected X-ray photons
ejected in the past by the central source Sgr A*. They also predicted that a flux of 6.4 keV K-$\alpha$ iron line
had to be observed in the direction of these clouds and that the continuum and the line emission had to be time variable
with a characteristic period needed for a photon front to cross the clouds. Later these effects were observed
by next generation X-ray  telescopes such as, ASCA and Suzaku \citep[][]{koya1,nobu11,ryu},
INTEGRAL \citep{Revnivtsev04,terrier}, Chandra \citep{clavel13}, XMM-Newton and NuSTAR \citep{ponti10,clavel14,zhang,kriv17}.
All these observations
can be perfectly described by introducing several X-ray flares emitted by Sgr A* in the past.

It is difficult to reproduce this phenomena using charged particle models. Indeed, charged particles are scattered by
interstellar turbulence and their propagation resembles diffusion. Therefore even if the source of the particles is transient,
characteristic emission time is determined by the longer one of the two timescales: (i) their propagation time from
the source to the emitting cloud, and (ii) their life time in the medium due to energy losses. Since all phenomena mentioned above
are characterized by a rapid temporal variability of the emission with timescale of the order of several years,
it is obvious that protons with very long lifetime can be safely ruled out.

Subrelativistic electrons responsible for the X-ray emission, on the other hand, are subject to very intense energy losses.
Therefore they potentially can reproduce observed temporal variations of the emission \citep{yusef}.
However to do so it is necessary to assume that there are several transient sources of electrons
located near the X-ray emitting clouds \citep{dog14}. This situation is not impossible but exceptional.
Indeed, in the case of X-ray reflection scenario, we need to set the temporal characteristics of the source of the flares only,
while for scenario with electrons we need to assume temporal and spatial position for the sources of electrons.
Therefore models with subrelativistic electrons could be considered as less viable.

However, the question arises whether the level of the continuum and the 6.4 keV line emission drops to zero
when the photon front has left a cloud or there is a background emission generated by any other process.
This emission can also be produced by bremsstrahlung of cosmic rays (CRs) and by K-$\alpha$ vacancy production in
iron atoms by subrelativistic electrons or protons \citep[see e.g.][]{dog98, tati03}.
Attempts to interpret generation of the continuum and line emission from the clouds by CRs were undertaken
in several models \citep[see e.g.][]{yus02,yusef,dog09,dog11,tat12}. However, the clearly observed time variability of X-ray fluxes
from the clouds was completely unfavorable to these models \citep[see e.g.][]{dog14}.

Although subrelativistic CRs (unlike relativistic CRs) do not produce visible radiation fluxes that could be detected in the Galaxy,
there are indications that their density is not zero in the interstellar medium.
Thus, the observed ionization of interstellar hydrogen may be produced by subrelativistic CRs \citep[see][]{indri12, dog13},
and the estimated energy density in the central molecular zone (CMZ) region could be
as high as 100 eV cm$^{-3}$ \citep[see][]{yus07,tat12,dog15}. Therefore a nonzero flux of X-rays from the clouds is expected
when the front of primary photons has left it. Attempts to estimate this flux from the cloud Sgr B2 were undertaken in \citet{dog14, dog15}.
According to their result the present 6.4 keV flux from this cloud is close to the expected stationary level,
but this estimates cannot be considered as reliable. There is no confirmation from observations that the Sgr B2 6.4 keV flux
has reached its stationary minimum, although its value has decreased by more than one order of magnitude from the peak.
Observations by \citet{zhang} suggested a possibility that stationary component started to appear, but reliable results
from further observations are necessary.

Interesting effects of X-ray variability were recently  observed in the direction of Arches cluster by \citet{kriv17}.
The Arches cluster is a cluster of young massive stars in the Galactic center.
It is likely associated with the ``$-30$ km s$^{-1}$'' molecular cloud. The mass of the cloud is estimated as $\sim 6\times 10^4~M_\odot$,
the hydrogen density there is about $n_H\simeq 10^4 $cm$^{-3}$, the gas column density $N_{H_2}\simeq 4 \times 10^{23}$ cm$^{-2}$
and the radius about 3 pc \citep{ser1987}. Continuum and 6.4 keV line X-ray emission was found
in the direction of the cloud by \citet{yus02,wang06}.
The line emission from the cloud varies with time \citep[][]{clavel14}
and is also characterized by a relatively high equivalent width,
therefore its origin is most likely due to the reflection of primary photons emitted by an external source.

Recently \citet{kriv17} found time variations of the continuum in the range 2-10 keV and the line 6.4 keV emission.
The essential result of this observation is that they found also  time variations of the line equivalent width, $eW$,
from $0.9\pm 0.1$ keV in 2007, to $0.6-0.7$ keV in 2015. This means that the Arches emission is a mixture of two components
with different equivalent widths. For sure, the time variable component observed by \citet{kriv17} can be interpreted as the
Thomson scattering of primary photons e.g. from Sgr A* which are leaving this complex. The second component can be due to either
the Thomson scattering of photons from another molecular complex which is at large enough distance from Arches but exactly
on the same path of view or due to a contribution of CRs into the total X-ray flux. In both cases the effect of $eW$ time variability
is naturally expected. Below we discuss both interpretations.

\section{Input parameters of the X-ray emission for the stationary and time-varying components}\label{seq:inputs}

As follows from \citet{kriv17}, the flux of 6.4 keV line was constant for the period 2002 - 2007 and
equals $I_0=8.84$ in units of $10^{-6}$ ph s$^{-1}$ cm$^{-2}$. After 2007 this flux  was decaying
with the rate $\alpha=0.64$ yr$^{-1}$ in the same units.

We assume that this flux consists of two components: a time variable component $C_{XR}$ produced by primary photons
from an external source which decays with time when $t\geq t_0$ i.e when the front of primary photons leaving the cloud,
and a stationary component $C_{C2}$ of unknown origin. We expect that at unknown time $t_X$ this flux has reached
the background stationary level of the 6.4 keV flux $C_{C2}$ when the front of primary photons left the cloud.
These temporal variations can be presented from \citet{kriv17} as
\begin{equation}
\frac{I_{6.4}(t)}{10^{-6}\mbox{ph}~\cdot\mbox{s}^{-1}\mbox{cm}^{-2}} =
\left\{
\begin{array}{ll}
C_{C2} + C_{XR}  & \mbox{if }t \leq t_0\\
C_{C2} + C_{XR}  - \alpha(t - t_0) & \mbox{if } t_0 < t < t_X \\
C_{C2} & \mbox{if } t \geq t_X
\end{array}
\right.
\label{eq:64time}
\end{equation}
where  $t_0 \approx 2007.4$ yr and $C_{XR}$ is the unknown contribution of varying X-ray component into the 6.4 keV flux from Arches.
The unknown time $t_X$  can be estimated as $t_X = t_0 + C_{XR}\alpha^{-1}$, if $C_{XR}$ is known.

It is natural to assume that the continuum emission $I_X(E_X)$ produced by primary X-ray photons evolves in the same way as 6.4 keV line emission.
We assume also that each of these components is characterized by  the different equivalent width of the 6.4 keV line,
$eW_{XR}$ and $eW_{C2}$, which do not equal each other, $eW_{XR}\neq eW_{C2}$ and remain constant in time. Here
\begin{equation}
eW=\frac{I_{6.4}}{I_X(E_X-6.4~keV)}\label{eW}\,.
\end{equation}

Then the total $eW$ in the time interval  $t_0 < t < t_X$ can be presented  as
\begin{equation}
eW = \frac{C_{C2} + C_{XR}  - \alpha(t - t_0)}{eW_{XR}^{-1}[C_{XR} - \alpha(t - t_0)] + eW_{C2}^{-1}C_{C2}} \,.
\label{eq:system_for_xr}
\end{equation}

From \citet{kriv17} we have: $eW(t = t_0) = eW_0 = 0.9 \pm 0.1$ keV, while at $t_1 = t_0 + \Delta t$,
where $\Delta t = 9$ yrs, $eW(t = t_1) = eW_1 = 0.65 \pm 0.06$ keV. Then Equation (\ref{eq:system_for_xr}) gives
\begin{eqnarray}
eW_{XR} &&= \frac{\alpha(t_1 - t_0)}{eW_0^{-1}(C_{XR} + C_{C2}) - eW_1^{-1}(C_{XR} + C_{C2} - \alpha\Delta t)}
  \nonumber \\
eW_{XR} &&= \frac{I_{6.4}(t_0) - I_{6.4}(t_0+\Delta t)}{eW_0^{-1}I_{6.4}(t_0) - eW_1^{-1}I_{6.4}(t_0+\Delta t)} = 1.1 \pm 0.3~\mbox{keV} \,.
\label{EW}
\end{eqnarray}

Two conclusions follow from this result. For solar abundance the equivalent width of the line generated by photons is \citep[see][]{tsuj07}
\begin{equation}
eW_{XR}^{\rm solar} \approx 3 \left(\frac{1}{\Gamma + 2}\right)\left(\frac{6.4}{7.1}\right)^\Gamma\left(\frac{1}{1 + \cos^2 \theta}\right)~\mbox{keV}\,,
\label{eq:EW_XR_pars}
\end{equation}
where $\Gamma$ is the spectral index of primary photons and $\theta$ is the reflection angle.
For estimates we take $\theta \approx \frac{\pi}{2}$. From this equation one can obtain that $eW_{XR}^{\rm solar} = 0.7$ keV
for the spectrum of X-ray continuum $\Gamma=1.6$ derived by \citet{kriv17}. Then from Equation (\ref{EW})
it follows that $eW_{XR}=\eta eW_{XR}^{\rm solar}$ ($\eta$ is the iron abundance relative to the Sun)
and it gives $\eta=1.6\pm 0.4$ in Arches, which is the same as derived by \citet{tat12}.

The other conclusion is that $eW_{XR}>eW_0$ i.e. even at times $t\leq t_0$ the contribution of stationary component into the total continuum
and line fluxes is nonzero.

As the next step we try to estimate  the contribution of stationary component $C_{C2}$ accepting that
the spectral index of the variable X-ray components equals $\Gamma=1.6$ and does not change in time.
For the given spectral index $\Gamma$,  $eW_{C2}$ cannot exceed $eW_1 = 0.65 \pm 0.06$ keV that gives
the upper limit for $C_{C2}$, $C_{C2}<3.06$

Then the 6.4 keV line intensity can be presented  from Eqs. (\ref{eW}) and (\ref{eq:system_for_xr}) as
\begin{equation}
eW_{C2} = \frac{eW_0C_{C2}}{I_{6.4}(t_0) - \frac{eW_0}{eW_{XR}}[I_{6.4}(t_0) - C_{C2}]} \,.
\end{equation}

\begin{figure}[ht]
\begin{center}
\includegraphics[width=0.8\textwidth]{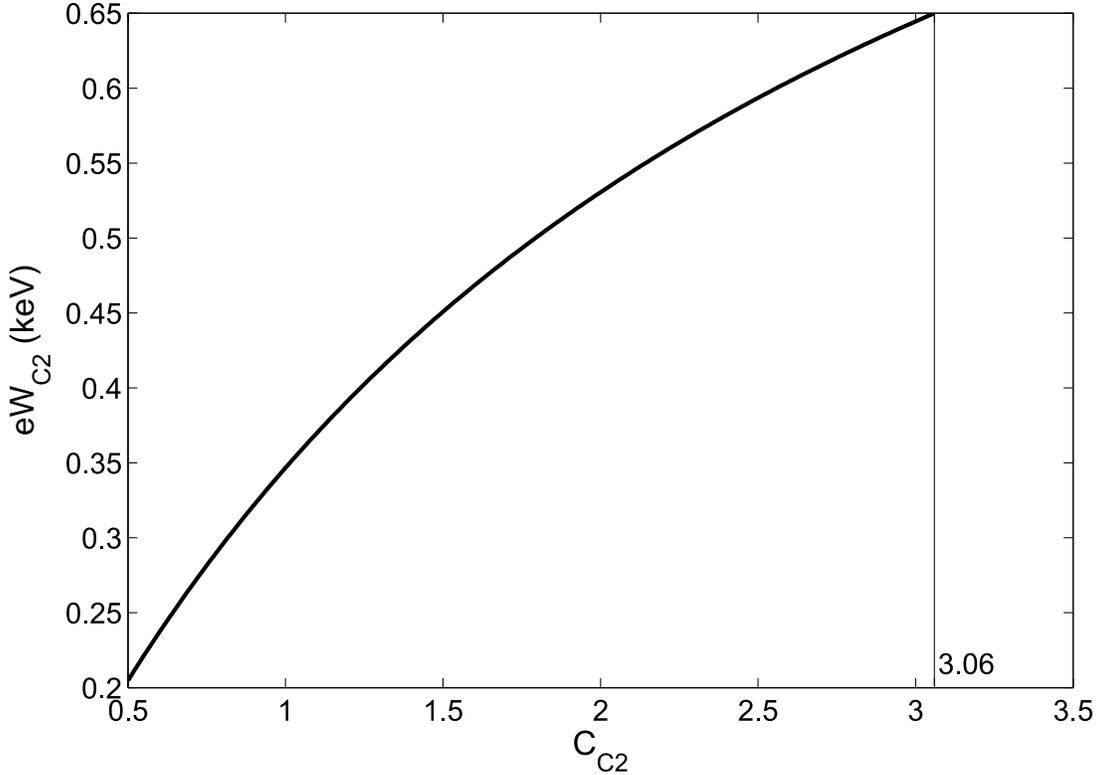}
\end{center}
\caption{Equivalent width of the stationary component as a function of its normalization.
Here Arches iron abundance equals $1.6$ of solar and $C_{C2}<3.06$.}
\label{fig:ccrew}
\end{figure}

The function $eW_{CR}(C_{C2})$ is shown in Figure \ref{fig:ccrew}. As one can see, equivalent width can be as high as 0.65 keV for $C_{C2}=3.06$.

Since the abundance of iron in Arches cluster is not fixed well, for theoretical estimations
we use the solar abundance of iron $n_{Fe}/n_H = 3\times 10^{-5}$ \citep{tsuj07} as a reference value.
The corresponding values of equivalent width for iron abundance of 1.6 solar can be obtained by multiplication of
solar abundance value by a factor of 1.6. For example, in order to reproduce aforementioned equivalent width of 0.63 keV
in Arches environment we need to obtain a value  about 0.3-0.5 keV for the solar abundance.

We notice that \cite{tat12}, using high-quality XMM-Newton data set, measured spectral index of the X-ray emission
from the Arches cloud as $\Gamma_X=1.6_{-0.2}^{+0.3}$, later confirmed by \cite{kriv14} with NuSTAR in 2012 ($\Gamma_X=1.6\pm0.3$).
Subsequent NuSTAR observations of the Arches cluster complex showed the trend of softening of the non-thermal power-law continuum.
\citet{kriv17} determined $\Gamma_X\sim2$ in 2015 observations, and a recent analysis of the NuSTAR data acquired in 2016
showed $\Gamma_X=2.7\pm0.5$ (Kuznetsova et al., in prep.), however the uncertainties are large.
Summarizing the above, we define that allowed values of the spectral index, following from observations, are within the limits
\begin{equation}\label{gx}
1\leq \Gamma_X\leq 2\,.
\end{equation}
We notice also that these results permit time variability of $\Gamma_X$ within the limits during the period of observations.

\section{Parameters of of the X-ray emission created by an additional X-ray flare}

As we mentioned above, $eW$ of the reflected X-ray emission depends on the following parameters:
the spectral index of primary flare, reflection angle and abundance of iron (see Equation (\ref{eq:EW_XR_pars})).

The most straightforward way to interpret these variations is to assume changes of the iron abundance in the complex
when the front moving along it. However since X-ray emission does not show significant spatial offset \citep{kriv17},
we find it highly unlikely that two components of the same complex have completely  different chemical compositions.

\begin{figure}[ht]
\begin{center}
\includegraphics[width=0.8\textwidth]{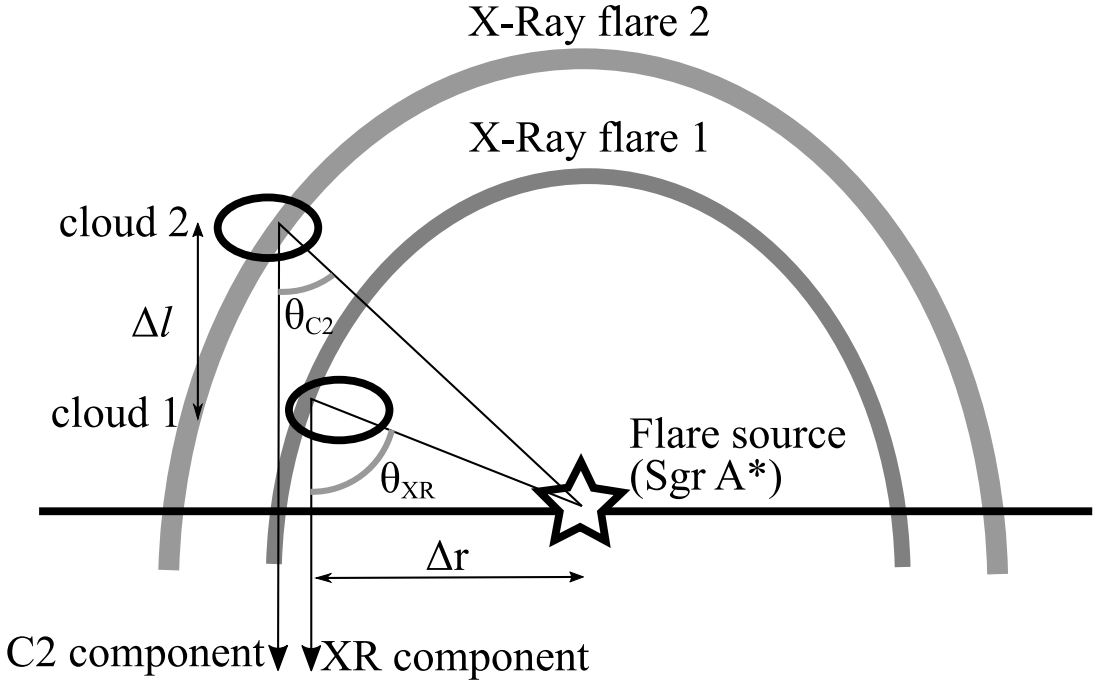}
\end{center}
\caption{Positions of the two clouds and possible reflection geometry.}
\label{fig:geom}
\end{figure}

The other way of explanation is to assume  that two separated molecular complexes are irradiated by the same
or by two different X-ray flares of an external X-ray source when the front of X-rays is inside the cloud during
the whole time of observation that provides the stationary component $C_{C2}$, while this (or the other) front is leaving
the second cloud that provides the time-variable component $C_{XR}$.
If the spatial separation between these clumps is large enough they are irradiated at different reflection angles,
$\theta_{C2}$ and $\theta_{XR}$, that mimic temporal variability of the total equivalent width $eW$.
Possible reflection geometry is shown in Figure \ref{fig:geom}.

In the Thomson regime one can relate the reflection angles in the following way:
\begin{equation}
\cos^2 \theta_{C2} = \frac{eW_{XR}}{eW_{C2}}\left(1 + \cos^2 \theta_{XR}\right) - 1 \,,
\end{equation}
The right-hand side of the equation is smaller than unity, that gives $eW_{C2}< 0.5eW_{XR}$. From Figure \ref{fig:ccrew} we conclude that $C_{C2} \geq 2$.

If primary photons are generated by the same external source then the separation distance between the complexes along the line of sight
$\Delta l$ can be presented as
\begin{equation}
\Delta l = \Delta r \left\{ \left[\frac{2 - \frac{eW_{XR}}{eW_{C2}}\left(1 + \cos^2 \theta_{XR}\right)}{\frac{eW_{XR}}{eW_{C2}}
\left(1 + \cos^2 \theta_{XR}\right) - 1}\right]^{-0.5} - \tan^{-1} \theta_{XR}\right\} \,,
\label{eq:separation}
\end{equation}
where $\Delta r = 25$ pc \citep[see, e.g.][]{yus02} is the projected distance between source of the X-ray flare
(hereafter we assume it is Sgr A$^*$) and emitting cloud. Here we assume that both components are located
either closer to us than Sgr A$^*$ or behind it. If $\frac{eW_{XR}}{eW_{C2}} \approx 1.52$, the spatial separation
reaches the minimum value of $\Delta l = 25$ pc for $\theta_{XR} \approx 0.45\pi$. Since the size of the Arches complex is of about $6$ pc,
the reflection regions belong to different molecular clouds.

From the equation for temporal delay $\Delta t_{XR}$ of the Compton echo for the $XR$ component \citep[see e.g.][]{sun98}
\begin{equation}
\frac{\Delta l}{c}=\frac{1}{2\Delta t_{XR}}\left[\Delta t_{XR}^2-\left(\frac{\Delta r}{c}\right)^2\right]
\end{equation}
that gives for $\theta_{XR} \approx 0.45\pi$  the temporal delay  about $\Delta t_{XR} = 100$ yrs which is in good agreement
with findings of \citet{chur17} and \citet{clavel17}, who estimated the age of the X-ray flare to be of about of 110 yrs.

The second component should have a reflection angle of $\theta_{C2} = 0.23\pi$ and temporal delay in this case is
about $\Delta t_{CR} = 230$ yr. It is highly unlikely that both of these components are created by the same flare,
since according to \citet{chur17} duration of the $110$ yrs old one is of order of several years.
It is more reasonable to assume for this scenario that there are two clouds located on the same line of sight
which are irradiated by two successive X-ray flares separated by about $130$ yrs. We note that timings of these two flares
are in very good agreement with so-called two-event scenario for X-ray emission from the Galactic center \citep{clavel13, walls16, terrier17}.
Just similar scenario was derived by \citet{clavel17} from the XMM-Newton and Chandra data,
who found from their analysis two flares of Sgr A*: 110 and 240 years ago.
According to \citep{clavel13} the older flare should be at least several decades long.
Therefore, this flare may be responsible for the stationary component $C_{C2}$.

Above we assumed that the Arches cloud is located further away than Sgr A*.
If it is located closer to us, temporal delays for given reflection angles should be the following:
$\Delta t_{XR} = 74$ yrs and $\Delta t_{C2} = 33$ yrs. The value of $\Delta t_{C2}$ is too low so we consider this situation as unlikely.

Relative positions of the clouds can be roughly estimated based on their absorption column density $N_{H_2}$.
Observations indicate that there is a slight decrease of absorption column density with time:
it drops from $7\times 10^{22}$ cm$^{-2}$ \citep{clavel14} to slightly below $7\times 10^{22}$ cm$^{-2}$ \citep{kriv17}.
This implies that the cloud responsible for component $C2$ absorbed less than the first one.
This may indicate that the second cloud is actually located closer to us,
but we do not really know what fraction of $N_{H_2}$ came from absorption in the medium located near the cloud
or within the cloud itself (i.e., a local effect) and not related to actual distance.
Indeed for other molecular clouds in the Galactic center described by \citet{ponti10} the absorption column density is
$4\times 10^{22}$ cm$^{-2} \leq N_{H_2} \leq 10\times 10^{22}$ cm$^{-2}$. Arches cloud is within this range.

The total luminosity of the 230 yr old flare in energy range 1-10 keV can be estimated as \citep{sun98}
\begin{equation}
L^X_{C2} \approx 6\times 10^{38}~\mbox{erg} \mbox{s}^{-1}~\left(\frac{C_{C2}}{3}\right)
\left(\frac{M_{C2}}{10^3M_\odot}\right)^{-1}\left(\frac{\eta}{1.6}\right)^{-1} \,,
\end{equation}
where $M_{C2}$ is mass of the cloud irradiated by the flare, $M_\odot$ is the solar mass and $\eta$ is the iron abundance relative to the solar.
Luminosity of the flare necessary to illuminate Sgr B2 is $L^X_{SgrB2} \approx 10^{39}$ erg s$^{-1}$  \citep{koya1}.
One can see that the assumption of the 230 yrs old flare is indeed the same one that illuminates Sgr B2 can be justified
for reasonable values of mass of the clump. Indeed, given the size of emitting region of about $a \times b = 1$ pc $\times ~2.3$ pc,
and absorption column density $N_{H_2} = 7 \times 10^{22}$ cm$^{-2}$ \citep{kriv17},
one can estimate an upper limit of the total irradiated mass as
\begin{equation}
M_{C2} \leq \pi a b \cdot m_p\cdot N_{H_2} \approx 3.7\times 10^3 M_\odot \,,
\end{equation}
and therefore each of the two components passes through $1.8\times 10^3 M_\odot$ of molecular gas.
Note that actual sizes and actual values of $N_{H_2}$ related to clouds should be different.
Therefore the value obtained above should only be used as a rough estimate.

The only difficulty of this model is that this scenario requires coincidence of very specific conditions
when the two complexes are exactly on the line of sight and they are separated exactly by a distance 25 pc.

\section{Spectral parameters of the X-ray emission created by charged particles}

\subsection{Basic equations for the emission created by charged particles} \label{seq:basic_eq}

We assume that the spectrum of primary charged particles (protons or electrons) inside the molecular cloud is a power law. Besides,
Fermi-LAT have not found  a gamma-ray flux from Arches at the level above  $10^{-5}$ ph cm$^{-2}$ s$^{-1}$.
It means that there should be significant steepening  in the the relativistic energy range of CR spectrum associated with Arches cluster.
The simplest way  is to introduce an effective cut-off at the  energy $E_{max}$ which is below the  limit of 100 MeV photon production.
For protons we can set it below the threshold of $p-p$ reaction, i.e $E_{max} \leq 200-300$ MeV, while for electrons
we can take $E_{max} \leq 100$ MeV. Then for the spectrum of CRs penetrating from outside we take it in the form
\begin{equation}
f_{p,e}(E) = AE^\delta \cdot \Theta (E_{max} - E) \,,
\label{eq:distr_function}
\end{equation}
where $A$ is a normalization constant, $\Theta(E)$ is the Heaviside function, the spectral index $\delta$
and the maximum energy $E_{max}$ are parameters of the model.

CRs penetrating from outside generate secondary electrons in the Arches cloud which also contribute to the total X-ray flux from there.
 For the stationary model the spectrum of secondary electrons can be derived from
\begin{equation}
\frac{\partial }{\partial E}\left( \frac{dE}{dt} f_{se}\right) = Q_{se}(E) \,,
\label{eq:secondary_equation}
\end{equation}
where $f_{se}(E)$ is the volume-averaged distribution function of secondary particles, $\frac{dE}{dt}$ describes energy losses
by ionization and bremsstrahlung \citep{blu70} and the term $Q_{se}(E)$ describes production spectrum of secondary particles,
\begin{equation}
Q_{se}(E) = n\int dE_p f_p(E_p) v \left(\frac{d\sigma(E_p,E)}{dE}\right)_{se}\,.
\label{Qesec}
\end{equation}
Here $(d\sigma/dE)_{se}$ is the cross-section of electron production by the knock-on process \citep{haya}.
Electrons produced by proton-proton collisions can be safely ignored since we only consider protons
with energies below the threshold of pion production. The rate of energy losses $\frac{dE}{dt}$ and the production function of electrons,
$Q_{se}(E)$ in Equation (\ref{eq:secondary_equation}) are proportional to the ambient density $n$. Therefore,
the resulting spectrum of secondary electrons, $f_{se}$ is independent of this medium parameter.
We ignore escape of secondary electrons from the cloud because their lifetime inside the dense cloud is quite short.

For the known distribution functions of protons, $f_p(E)$, and electrons, $f_e(E)$ and $f_{se}(E)$,
the spectrum of the X-ray continuum can be estimated as
\begin{equation}
I_x(E_x) = n \sum_{p,e} \int dE ~f_{p,e} v \left(\frac{d\sigma(E,E_x)}{dE_x}\right)_{br} \,,
\label{eq:continuum_CR}
\end{equation}
where $\left(\frac{d\sigma(E,E_x)}{dE_x}\right)_{br}$ is the cross-section of inverse-bremsstrahlung emission for protons
and of bremsstrahlung emission for leptons \citep{blu70}. We take also into account electron-electron bremsstrahlung \citep{haug}
whose contribution is significant  for hard spectra of electrons. Indeed as one can see from \citet{haug} 10 MeV electrons
produce about 1.5 times more of 6.4 keV photons through electron-electron in comparison to electron-proton bremsstrahlung.

The intensity of 6.4 keV line is estimated as
\begin{equation}
I_{6.4} = n\eta \sum_{p,e} \int dE ~f_{p,e}  v \sigma^{K\alpha}_{Fe} \,,
\label{eq:Feline_CR}
\end{equation}
where $\eta$ is a relative abundance of the iron atoms on the cloud and $\sigma^{K\alpha}_{Fe}$ is a cross-section of
production of 6.4 keV photon by proton and electron impact \citep[see][]{tati03}.

When the intensity of the X-ray emission is known, it is possible to estimate the total power in charged particles (protons or electrons)
required to produce this emission. The difference between protons or electrons is about of a few since their cross-sections
for producing X-ray photons are are of the same order \citep{tati03}. The luminosity of the source of charged particles
estimated by \citet{tat12} is about $\sim 10^{39}$ erg s$^{-1}$.

Another important parameter of X-ray emission following from observation is the steepening of its spectrum above 10 keV.
According to \citet{kriv17}, the X-ray spectrum is steepening from $\Gamma_X = 1.6$ at $E_X<10$ keV to $\Gamma_X = 2$ at higher energies.
This spectral change can be reproduced by introducing a spectral break $E_{br} \approx 200$ MeV for protons
(that is about the required value of $E_{max}$) and at $E_{br} \approx 100$ keV for electrons.

\subsection{Pure hadronic and leptonic models with stationary spectral index}

\begin{figure}[ht]
\begin{center}
\includegraphics[width=0.8\textwidth]{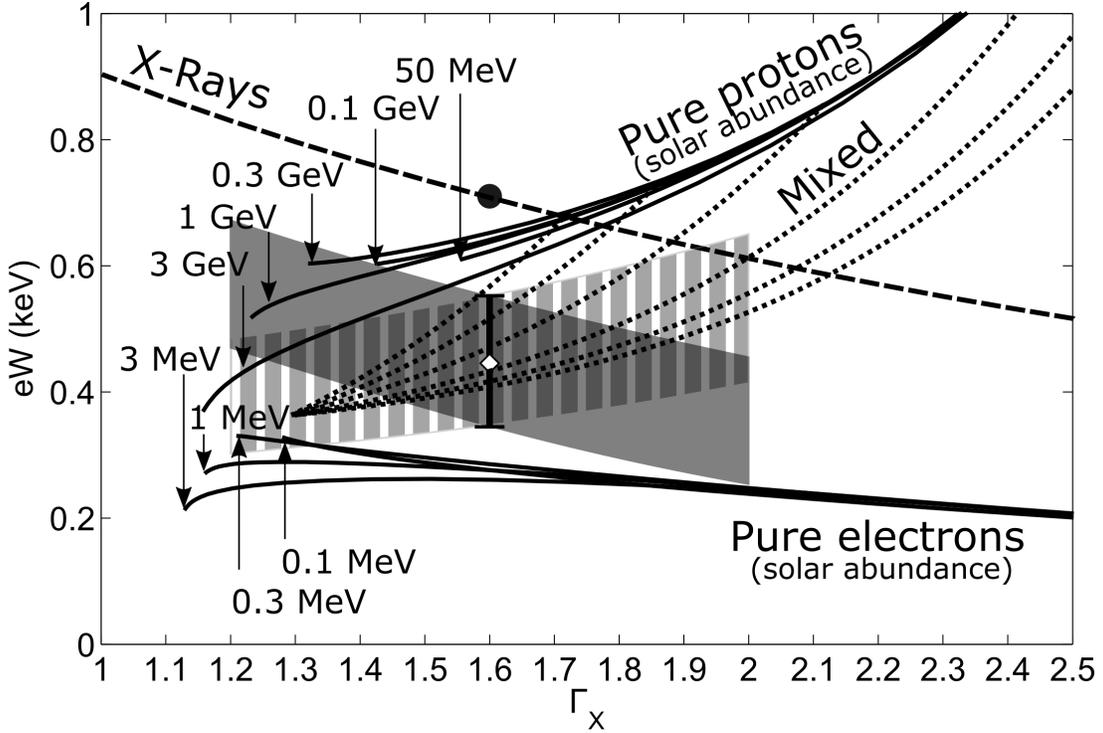}
\end{center}
\caption{Equivalent width as a function of spectral index of the X-ray emission produced by charged particles or photons
in the energy range between 1 keV - 7 keV. Solid lines correspond to pure hadronic model (top curves) and pure leptonic models (bottom curves)
with different values of maximum energy $E_{max}$. Dashed line corresponds to the emission produced by primary X-rays.
Dotted curves correspond to mixed (hadronic plus leptonic) models with different fractions of components.
Data points in the middle - equivalent width at 2015 calculated in Section \ref{seq:inputs}.
Gray area - possible variations of the experimental value of equivalent width in the case
if spectral index of the emission is different from $\Gamma = 1.6$: solid area -
if spectral index of the X-ray emission does not vary with time,  striped area -
if spectral index of the X-ray emission is not constant. Black dot corresponds to the time-varying component prior to 2007.
Solar abundance of iron is assumed.}
\label{fig:ew}
\end{figure}

Here we assume that the spectral index of X-ray emission produced by CRs (stationary component) equals exactly
that of X-ray emission generated by an external source (variable component).
For the allowed range of $\Gamma_X$ (see Equation (\ref{gx})) one can estimate from Equations (\ref{eq:continuum_CR})
\& (\ref{eq:Feline_CR}) the equivalent width of the 6.4 keV line $eW_{C2}$ generated by CRs. The result $\eta=1$ is shown
in Figure \ref{fig:ew} by the solid lines where we plotted equivalent width as a function of spectral index of  X-ray emission $\Gamma_X$
between 1 keV and 7 keV. The curves at the top right corner show the function $eW(\Gamma_X)$
for protons and curves at the bottom that for electrons. These curves define the background level of $eW$ for the case of protons or electrons,
when the X-ray front has left the cloud. Each curve corresponds to different $E_{max}$:
we used values $50$ MeV $\leq E \leq 3$ GeV for protons and we used values $100$ keV $\leq E \leq 3$ MeV for electrons.
One can see that equivalent width weakly depends on the value of $E_{max}$.
The minimum allowed value of $\Gamma_X$ however depends on $E_{max}$, and therefore it is unlikely to generate X-ray emission
with hard spectrum by protons ($\Gamma_X < 1.6$) if their maximum energy is below $E_{max} < 40$ MeV.

Curves with high values of $E_{max}$ are included in Figure \ref{fig:ew} for the sake of generality.
Indeed, as we mentioned in previous section, relativistic particles should generate a prominent flux in gamma-rays.
For example, for $E_{max} = 3$ GeV and the minimum value of $\Gamma_X$ corresponding to $\delta = 0.5$,
the gamma-ray emission above 100 MeV is expected to be of the order of $10^{-5}$ ph cm$^{-2}$ s$^{-1}$.
Also steepening of the X-ray spectrum  above 10 keV cannot be reproduced if the spectral index of primary particles remains constant.

In Figure \ref{fig:ew} we also plotted equivalent width of the 6.4 keV line produced by primary X-rays.
The corresponding dependence is shown by dashed line and it represents the time-variable ($C_{XR}$) component of
the emission observer before 2007. The value $eW$ derived from \nustar~ data for $\Gamma_X = 1.6$,
is shown by the data points in the middle of the figure. The estimated values of $eW$ for the range of $\Gamma_X$ defined by Equation (\ref{gx})
if it is time independent is shown in the gray area of Figure \ref{fig:ew}.

Regardless of the emission process the value of $eW$ is proportional to the abundance of iron.
Therefore despite absolute positions of the curves in Figure \ref{fig:ew} do depend on the iron abundance,
their relative to each other positions do not. With this in mind we plotted all curves in Figure \ref{fig:ew}
assuming solar abundance of iron for convenience.

One can see that although protons can provide a drop of the equivalent width with time, the magnitude of this drop is not big enough
to reproduce the results of \cite{kriv17} except for the case of $E_{max} \geq 0.1$ GeV when the spectral index $\Gamma_X < 1.4$.
Then the theoretical curves fall into the shaded area, defining the background level produced by protons.
As follows from observations of \citet{tsuj07} the X-ray continuum spectrum in Arches is indeed very hard with index of $\Gamma_X < 1.4$,
yet background contamination is possible. In addition, for $E_{max} \leq 0.2$ GeV spectral turn-over at $E_x=10$ keV,
reported by \citet{kriv17}, is also reproduced. From Figure \ref{fig:ew} one can see that in the case of pure hadronic model,
the X-ray emission in 2015 has reached its background (stationary) level and is fully produced by protons,
i.e., the X-ray front has left completely Arches area.

We notice that the scenario of bombardment by low-energy protons with hard spectra may interpret the 6.4 keV production
in the Sgr B2 molecular cloud \citep{dog15} and in the Inner Galactic Ridge \citep{nobu15}.

For pure leptonic model the situation is different. Electrons are characterized by very low equivalent width.
According to Figure \ref{fig:ccrew} the contribution of electrons to the total X-ray flux in pure leptonic model should be low,
and appearance of the stationary component expected in several years as it follows from Figure \ref{fig:ew}, if it can be measured.

The problem of the electron model is their short lifetime. Lifetime of 10 keV electrons in the medium with density
of $10^4$ cm$^{-3}$ is less than 0.1 yr. Penetrating from outside they can fill the shell with a thickness about $10^{16}$ cm unless
they are accelerated inside in-situ. However, if the energy of primary electrons is about 1 MeV they can fill the whole volume of the cloud.

Spectrum of X-ray emission produced by electrons penetrating into molecular clouds was analyzed by \citet{tat12}.
They assumed that according to \citet{skil76} low-energy electrons are excluded from the molecular clouds.
Therefore spectrum of low-energy electrons is formed entirely by energy losses and since at low energies ionization losses dominate,
the spectrum of electrons should be very hard. As a result, the index of X-ray spectrum produced by electrons should be low: $\Gamma_X \leq 1.4$

Moreover, in order to reproduce the spectral break at hard X-rays one need to assume that there is a cut-off or a hard spectral break
in the spectrum of primary electrons at $E_{max} \leq 100$ keV. The corresponding lifetime of these electrons is about 1.4 yrs.

Theoretical curve for electrons is located below the shaded area.
As it follows from Figure \ref{fig:ccrew} stationary level of pure leptonic model corresponds to $C_{C2} \approx 2$,
i.e., should be about 1.5 times lower than observed in 2015.

Therefore the main difference between pure hadronic and pure leptonic model is the value of stationary level:
in the case of pure hadronic model the stationary level have already been reached,
while in the case of pure leptonic model it is still below the current observations.
In both models the spectral index of the X-ray emission should be rather hard: $\Gamma_X \approx 1.3-1.4$,
which allows us to separate these CR models from the model of X-ray flares irradiating two different clouds
(two Compton echos with different equivalent widthes).

\subsection{Models with temporal variations of spectral index}

Above we analyzed the case when spectral indexes of continuum X-ray emission generated by photons and charged particles equal each other.
In this case the spectral index of total emission is independent of time even if relative contributions are time varying.
Such a situation is possible but exceptional. It is more natural to assume that spectral indexes of photon
and the CR component differ from each other. Then the spectral index of the total X-ray emission is a function of time, $\Gamma_X(t)$.

If we accept variations of the continuum index from $\Gamma_X=1.6$ in 2007 to $\Gamma_X\neq 1.6$ in 2015,
then values of $eW$ of 2015 should be re-calculated. To do this, we re-evaluated spectral model of the cloud emission
\citep[Model~2 in Table~4 in][]{kriv17} on XMM-Newton data acquired in 2015, to estimate $eW$ for fixed $\Gamma_X$ running
in range $1.2-2$ (with step of 0.1).
The expected changes of $eW$ for different values of $\Gamma_X$ are shown in the gray area of Figure \ref{fig:ew}.
One can see that curves that the proton scenario is completely unacceptable because their stationary level is higher than measured in 2015.

The situations for electrons as one can see from Figure \ref{fig:ew} is similar to that described in the previous section.
However, in this case the spectral index of the variable X-ray component and that of stationary component are different,
and the stationary level generated by electrons can be reached already now if the spectral index of bremsstrahlung emission
is about $\Gamma_X \approx 1.3-1.4$.

\subsection{Mixed models}

A more realistic model should include contribution from both protons and electrons.
Indeed, in shocked plasma as it was shown by \citet{bar00} contribution of inverse bremsstrahlung produced by protons
can be safely neglected \citep[see, however,][]{nobu18}. However as we move away from the source of cosmic rays,
energy losses suppress density of electrons and a relative contribution of hadronic emission increases.
Thus, we expect that both electrons and protons contribute non-zero fractions of the X-ray emission into the total stationary flux from the Arches.

We use the equations from Section \ref{seq:basic_eq} assuming the following spectra for particles. For electrons it is fully defined  by losses:
according to \citet{tat12}, if spectrum of low-energy electrons injected into the cloud is hard enough \citep[see, e.g.][]{skil76},
then inside the cloud the spectrum should satisfy the expression
\begin{equation}
f_e(E) = A \left( \frac{dE}{dt}\right)^{-1}\Theta(E_{max} - E) \,,
\label{eq:fe_mixed}
\end{equation}
where  $\frac{dE}{dt}$ describes energy losses experienced by electrons.
As we already mentioned before, energy losses of electrons are more severe at low energies.
Therefore spectrum of electrons obtained from Equation (\ref{eq:fe_mixed}) is hard.
For protons we use spectrum in the form described by Equation (\ref{eq:distr_function}).

We consider both scenarios for stationary and time-variable spectral indexes.
Equivalent width as a function of spectral index for different proportions of electrons and protons and for different spectral
indexes of protons are shown in Figure \ref{fig:ew} as dotted lines.
One can see that combination of hard spectrum produced by electrons and soft spectrum produced by protons can potentially
reproduce any observed equivalent width and spectral index.

Different positions on the dashed lines correspond to different ratios between electrons and protons which is varying along the curved.
Bottom-left end of the lines corresponds to pure leptonic models, while the point of intersection of the dashed lines
and the proton lines correspond to pure hadronic models.

\section{Discussion and conclusion}

We investigated two scenarios to reproduce the observed variations of equivalent width of iron K$\alpha$ line
observed from the direction of the Arches cluster. We assume that there is a different component of the X-ray emission
which varies much slower and therefore can be considered as stationary. The second component can be produced either by a different
primary X-ray flare or by subrelativistic cosmic rays.

The most conservative way to explain variations of equivalent width of iron K$\alpha$ line is to use a two-event model proposed
by \citet{clavel13} and recently updated by \citet{clavel17}.
There are two flares in this model occurred in the GC: one about 100-yr old and the other about 200-yr old.
They irradiated two different clouds on the light of view, separated by a distance  $\ga 25$ pc.
Because of different reflection scattering irradiated emission from this clouds is characterized by different equivalent widths.
The reflection angle of the older flare is smaller than that of younger one.
Therefore it has a smaller equivalent width. As it follows from results of Section 2,
we should assume that this flare is responsible for the stationary component of emission
and hence the duration of this flares should be long enough.

For the timings of the flares taken from \citet{clavel17} the second component from the older flare should already be observed
and therefore stationary level should be already archived.
However if timing of the second flare is different, for example if it is older than the two reported by \citet{clavel17},
stationary level can be lower. However it will require more significant spatial separation between emitting clouds,
and the stationary level of the 6.4 keV line emission can be reduced only by a factor of 1.5 in comparison to the intensity observed in 2015.

Explaining the same variations by charged particles has some shortcomings but cannot be completely ruled out.
If we accept variations of X-ray spectral index within $1.3 \leq \Gamma_X \leq 2$, we arrive at the following conclusions:
\begin{itemize}
\item Pure hadronic model requires a stationary X-ray spectral index of the total emission, $\Gamma_X = 1.3$,
in the period from 2007 to 2015. Otherwise the value of the equivalent width of the iron line produced by protons
would be too large to reproduce the observed variations. As it follows from the hadronic model we have reached the stationary level
of the X-ray emission from Arches in 2015.
\item According to \citet{tat12} pure leptonic model requires a hard spectral index $\Gamma_X = 1.3$ when stationary level is attained.
However, unlike hadronic model, the stationary level of the 6.4 keV line emission may be 1.5 less than observed in 2015.
Therefore in 2015 we may still observe a combination of varying component produced by primary X-rays and a stationary component produced by electrons.
In this case spectral index of the emission can be softer: $\Gamma_X > 1.3$.
Future observations can potentially measure the spectral index more carefully and therefore restrict these models.
\item In the case of mixture of protons and electrons the value of spectral index is not really restricted.
Therefore it is quite difficult to distinguish between charged particles and X-ray.
Information on ionization and nuclear lines \citep{tat12} might be essential for the clarification.
\end{itemize}

The total CR power required to generate observed X-ray fluxes from Arches is about $4\times 10^{38} - 10^{39}$ erg s$^{-1}$,
which is close to the findings of \citet{tat12}. As it was showed by \citet{tat12}, this power can be generated by
collision between Arches cluster and Arches cloud.
Observations show that Arches cluster moves towards North-East in the equatorial coordinates \citep{stole08},
and regions bright in 6.4 keV line are located to the the North, East and South-East from the cluster \citep{kriv17}.
Therefore it is possible to assume that CR are accelerated near the sites of collision between Arches cluster and dense gas clumps.

In our analysis we assumed that the cloud responsible for time-varying component is located in the same plane as Sgr A*
and that the reflection angle is close to $\pi/2$. That resulted in relative iron abundance of about 1.6 solar value.
This assumption however may be incorrect, and the cloud in question may be located closer to us or further away,
resulting in higher iron abundance necessary to reproduce observed equivalent width.
This changes in application to charged particles models will proportionally shift the data point and the gray areas in Figure \ref{fig:ew} downward,
making electron scenarios more viable and proton scenario less viable.

X-ray reflection scenario is more sensitive to the assumed iron abundance.
Indeed as one can see from Equation (\ref{eq:separation}), the separation between the two clouds is very sensitive to
the reflection angle $\theta_{XR}$. If $\theta_{XR}$ tends to $0.3\pi$ and iron abundance tends to about 2 solar value,
the implied separation between the two clouds tends to infinity.
This information may significantly restrict or even rule out the X-ray reflection model, if iron abundance can be measured independently.

Despite the fact the all of the discussed models have limitations, we find that the X-ray reflection model is the most viable.
For X-ray model to work we only need to assume the specific positions of the irradiated clouds.
The strongest support is provided by the required timings of the flares for the model to work coincide with the values obtained independently.
In comparison, models involving charged particles require additional local particle accelerator located nearby the specific clouds.
Although there are some indications of interaction of Arches cluster with surrounding molecular gas, it is not clear if there is a shock
near the clouds bright in X-rays and if the shock in question is able to provide the necessary power to the charged particles.

However, there are some possible challenges to the X-ray reflection model:
\begin{itemize}
\item The model requires the clouds to be in specific locations.
Future observations on the relative distance between the two irradiated clouds will be crucial for the model.
\item The X-ray reflection model predicts that when the second flare leaves the cloud,
the intensity of the X-ray emission will start to decrease again.
Since we do not know the size of the cloud and the duration of the flare,
it is difficult to specify exact moment of time. However, given the fact that the duration of the flare is of order of 10 years,
this decrease may be observed in the near future.
In the frame of CR models, intensity of X-ray emission should stay constant within 0.5-1.0 of currently observed values.
\item As we already mentioned, X-ray reflection model and proton model are very sensitive to assumed iron abundance.
If independent observations show that iron abundance is higher than we used in our calculations,
only electron model will be able to reproduce the observed emission properties.
\end{itemize}

\acknowledgments
%\section*{Acknowledgments}
The authors are grateful to E.M. Churazov for useful suggestions. VAD and DOC are supported in parts by the grant RFBR 18-02-00075.
DOC is supported in parts by foundation for the advancement of theoretical physics and mathematics ``BASIS''.
RK acknowledges support from Russian Science Foundation (grant 14-22-00271).
CMK is supported in part by the Ministry of Science and Technology of Taiwan under grants MOST 104-2923-M-008-001-MY3 and MOST 105-2112-M-008-011-MY3.
KSC is supported by the GRF Grant under HKU 17310916.

\end{document}